# Success in Science:
# How Global Prestige Organizes Careers

**Marek Kwiek**
Center for Public Policy Studies, Adam Mickiewicz University, Poznan, Poland
kwiekm@amu.edu.pl, ORCID: orcid.org/0000-0001-7953-1063

**Wojciech Roszka**
(1) Poznan University of Economics and Business, Poznan, Poland
(2) Center for Public Policy Studies, Adam Mickiewicz University, Poznan, Poland
wojciech.roszka@ue.poznan.pl, ORCID: orcid.org/0000-0003-4383-3259

## Abstract

This article analyzes the structure of perceived academic success. We combine survey data from 10,848 Polish scientists with their Scopus bibliometric data at the individual level. We use polychoric correlations, exploratory factor analysis, network modeling (EBICglasso), and generalized linear mixed models in ordinal and binary forms. Our results show that academic success is multidimensional, with a clear core. This core is global publication prestige. Publishing in top international journals is the node with the highest centrality, and it is connected to other career dimensions, such as citations and international collaboration. Publications in top national journals, in contrast, are peripheral. The threshold structure of the scale indicates a selection effect. The definition of success is globally oriented and strongly tied to the hierarchy of international journals.

## 1. Introduction

Science is a strongly stratified system. Research in the sociology of science shows that publication productivity and scientific impact as measured by citations are extremely unevenly distributed (Cole & Cole, 1973; de Solla Price, 1963; Lotka, 1926; Merton, 1968). A small group of researchers produces a large share of publications and citations. They also obtain grants more often and receive more symbolic recognition. These inequalities are strengthened by mechanisms of cumulative advantage and strongly supported by institutional reward systems.

Stratification is increasingly powered by quantitative indicators applied by institutions to assess the research achievements of their faculty. Publication and citation numbers, as well as global journal rankings, increasingly matter at individual and institutional levels. They have all become basic tools for assessing research work. National funding and promotion systems rely heavily on bibliometric indicators, and this is especially true for measures linked to internationalization and publishing in prestigious journals. Success is usually defined by measurable outcomes and one's position in global science.

Despite the central role of metrics in academic life, we still know little about how scientists themselves understand academic success or how they order its components, if they do. Scientometric research widely analyzes objective indicators of achievement, but it less often discusses subjective definitions of success. Success is usually treated as an externally defined outcome measured using publications, citations, and research grants. It is less often treated as a multidimensional construct that is routinely interpreted by members of the academic community.

Contemporary academic science operates within specific evaluation regimes built around specific indicators (Scopus and WoS journal lists, national journal lists, etc.). Researchers' perceptions of success may show how they internalize these regimes. The options include, for example, the importance they assign to international publications, citations, research grants, academic awards and distinctions, institutional prestige, and formal promotions. The structure of perceived success can show how academic stratification works. It can also show whether the success system is a single hierarchy based on a single dominant principle, such as the prestige of international journals, or whether it is based on several partly autonomous logics of recognition.

In this study, we address this gap. We combine large-scale survey data with bibliometric data at the level of individual scientists. We analyze responses from more than 10,000 research-active scientists affiliated in Poland. We deterministically link our survey records to Scopus Author IDs and create full publication histories of all the respondents, from their first to their latest publication. This allows us to examine how the various dimensions of success are connected to each other. It also allows us to test how these assessments of success depend on individual, disciplinary, institutional, and bibliometric characteristics. We propose a relational view of success, and we focus on the links among the various dimensions of success.

We address three research questions:

1. Does academic success have one latent dimension, or does it have several partly independent dimensions?
2. Which aspects of success are central? In particular, does publication in top journals form the core of the system of success assessment?
3. Are the mechanisms that differentiate success assessments linear, or are they threshold-based? In other words, do threshold-based assessments appear when the highest, elite levels of achievement are defined?

To answer these questions, we use several analytical techniques. First, we estimate polychoric correlations for ten success dimensions measured with a Likert scale (the ten dimensions are listed in Figure 1). This approximates relationships among the latent continuous variables behind the responses. Next, we conduct exploratory factor analysis (EFA), using parallel analysis and an oblique rotation. This identifies the latent structure. We then reconstruct the relational architecture by estimating a partial correlation network with EBICglasso regularization. We also compute node centrality. Finally, we use multidimensional scaling (MDS), which shows the configuration of success dimensions in a two-dimensional space.

In addition to structural analyses, we use generalized linear mixed models (GLMMs). These models assess how individual characteristics influence the hierarchy of success assessments. In the models, we include such variables as gender, academic age, publication productivity, international collaboration rate, and intrinsic motivation in research work. We treat success as a coherent system of assessments rather than a set of independent outcomes. The ordinal cumulative logit model uses the full 5-point scale. We also estimate binary models (≥4 and =5, where 5 is the highest answer in the Likert scale). This allows us to test whether the determinants of "high" success differ from the determinants of "elite" success, testing threshold mechanisms.

Our results show that academic success is a hierarchically organized, multidimensional construct. Factor analysis reveals four distinct yet correlated dimensions: international publication visibility, institutional prestige and symbolic recognition, national achievements, and formal career advancement. Our network analyses show that publishing in top international journals is the central element in the network. It connects different aspects of perceived success. In contrast, the title of full professor and publications in top national journals are structurally peripheral. The global seems to matter more than the local in assessing what success is to Polish scientists.

The threshold structure of the Likert scale shows that the highest (elite) response category requires a stronger shift on the latent success dimension. Binary models confirm that some factors operate more strongly when elite (5) rather than very high (4) success is defined. This is especially true for the two variables of intrinsic motivation and institutional pressure. There are also differences between men and women: women assign the highest ratings to various success dimensions less often than men, which suggests more restrictive evaluation standards for men. Importantly, some bibliometric indicators do not shift the entire distribution of ratings evenly. They become important mainly at the highest levels of the scale.

The results therefore suggest that contemporary academic success is organized around international research visibility (see Leahey, 2007). Global journal prestige works as a key mechanism that organizes the system of assessments. At the same time, success remains internally differentiated. Academic careers cannot be reduced to one metric hierarchy. Success works as a threshold system: moving from high to elite success involves additional selection mechanisms.

Our contribution has three dimensions. First, empirically, we use a large, integrated dataset that links survey-based subjective success assessments with full bibliometric data at the individual level. Second, methodologically, we show the value of combining polychoric correlations, network analysis, MDS, and mixed models. And third, conceptually, we interpret success as a multidimensional, relational, and threshold construct, going beyond a view of success as a simple bibliometric outcome.

# 2. Theoretical framework and literature review

## 2.1. Academic success as part of a stratification system

Academic success can be analyzed in a broader context of inequality in science. Scientometric and sociological research has long shown that scientific productivity is highly concentrated (de Solla Price, 1963; Lotka, 1926). Publication and citation distributions are highly asymmetric, and a small share of researchers produces a large share of knowledge. We have shown the 10/50 rule in our earlier work on top research performers for Europe (11 countries) and for Poland: 10% of scientists produce 50% of publications (Kwiek, 2016; Kwiek & Roszka, 2024).

Classic analyses emphasize that such inequalities result from mechanisms of cumulative advantage, with earlier success increasing the chances of later success (Cole & Cole, 1973; Merton, 1968). Academic success in this account is not only an outcome of talent or hard work but also part of a reward system in science. Publications, citations, grants, and academy memberships signal the position a scientist is taking in the hierarchy. And the hierarchy is reproduced by institutional selection mechanisms. These mechanisms are durable over time (Kwiek & Szymula, 2025a). Research on scientific elites and productivity concentration confirms the stability of this structure (Allison & Stewart, 1974; Stephan, 2012), highlighting that academic success is both a normative and a stratifying category.

## 2.2. Journal prestige and global hierarchies of visibility

Publishing in high-impact journals signals quality (Heckman & Moktan, 2020) and increases the chances of future citations and access to more prestigious resources. Articles in prestigious journals can be converted into grants which, in turn, using equipment, staff, software etc., can be converted into new articles, following the “credibility cycle” in science (Latour & Woolgar, 1986).

The development of bibliometric indicators and journal rankings has further formalized prestige hierarchies. Impact Factor, CiteScore, and other indicators have become part of global and national evaluation infrastructures. This applies to both individual and institutional evaluations (Moed, 2005; Waltman, 2016). As a result, the prestige of international journals has started to function as a dominant principle that organizes academic careers, especially in highly internationalized disciplines (Fochler et al., 2016).

The global circulation of knowledge is structurally hierarchical. There are the most prestigious journals, less prestigious journals, and peripheral journals. Access to the most prestigious journals is unevenly distributed across regions and languages. This shapes patterns of visibility in science (Kwiek, 2021; Larivière et al., 2013; Sugimoto & Larivière, 2018). Therefore, publication success is strongly linked to a scientist’s position in the global hierarchy of scientific communication.

### 2.3. Success as a relational and threshold construct

This study treats success as a relational construct, without reducing it to one indicator. We analyze links between different dimensions of success, and we also examine their relative centrality in the system of assessments.

Threshold dynamics are also important for our interpretation. Moving from a high rating of a success variable to the highest rating may require additional conditions. These conditions may be more restrictive. As a result, the mechanisms that define elite success may differ from those that define high (and more accessible) success. This approach links a stratification perspective with an analysis of perception. Eliteness is not only an objective state but also an outcome of normative classification.

In the literature, success is often measured with productivity and citations. However, the literature emphasizes that success is multidimensional. In addition to publications, grants, awards, affiliation with elite institutions, and formal promotions matter (Abramo et al., 2017; Stephan, 2012). These dimensions do not have to overlap, and they may also follow different selection mechanisms.

We link survey data with full bibliometric microdata, which allows us to connect an analysis of the structure of success with an analysis of its determinants. We include individual and institutional factors to propose an approach in which academic success is a multidimensional, relational, and threshold system of assessments. This system is embedded in the global prestige hierarchy.

## 3. Data and methods

### 3.1. Large-scale survey of scientists

The survey questionnaire was developed based on literature on the academic profession. It was designed to collect data from scientists with at least one research article published in internationally visible journals (research-active scientists). These were scientists for whom Scopus provided a public email address in January 2023. The target population included all internationalized scientists and researchers affiliated in Poland. The survey was conducted on the Qualtrics platform from May to September 2023. Two reminders were sent in this period.

Invitations to participate were sent out to 65,300 scientists. Of these, 13,694 opened the survey. Full responses were obtained from 11,315 respondents. Another 226 respondents completed between 50% and 99% of the questionnaire. A further 2,153 respondents stopped after answering less than half of the questions. The sample used in this study includes 10,788 observations. Some variables have missing data, so the number of valid responses may differ in analyses, particularly in multivariate models where only complete cases are included. The sample covers respondents who indicated one of ten fields as their leading field. Because some had small counts, we aggregated the fields into five groups. The final response rate was 20.97%. This should be considered a good result for a nationwide, extensive survey with a median completion time of 40 minutes.

## 3.2. Data integration: survey and bibliometric data

The study uses data from two independent sources. The first source is the Scopus bibliometric database. The second source is the survey conducted with Polish academics. In this study, we follow the idea of an extended survey (Das & Emery, 2013; Salganik, 2018). The key that links the two datasets is the Scopus Author ID. For the problem of unique author identification, Scopus is considered more precise than Web of Science (Sugimoto & Larivière, 2018). We linked bibliometric data and survey data deterministically using the Scopus Author ID, which was known for all individuals invited to participate in the survey. We did not use probabilistic linkage, as we did in earlier work (see Kwiek & Roszka, 2021).

We also used the national administrative register of all Polish scientists, the RAD-on database. We used it to compare the structure of our sample with that of the population of scientists active in Poland in 2023. The comparison used selected characteristics: gender, age, academic position, and field of science. RAD-on covers all academic staff employed in all sectors of the Polish science system. We used it to check the representativeness of the survey.

We sent survey invitations to scientists for whom we had two types of data. First, we had a broad set of raw bibliometric data. This included, for example, the total number of publications and an individual four-year Field-Weighted Citation Impact (FWCI) indicator. Second, we had variables derived from further bibliometric processing. These variables included, among others, the dominant discipline based on the All Science Journal Classification (ASJC): to obtain a dominant discipline for every scientist in our sample, we used all cited references from all published articles, linking them to ASJC disciplines. We also had lifetime productivity data and prestige-normalized productivity data calculated using full counting mode. We included membership in selected classes of scientists (top performers and bottom performers, or the top 10% and bottom 10% in terms of productivity; internationalists and locals, that is, scientists with at least 50% of their articles published in international collaborations and the rest). We included institutional research intensity, distinguishing between the 10 IDUB universities (selected for a national research excellence program) and the rest. We also had measures of overall and international collaboration intensity. We had the year of first publication, which allowed us to calculate academic age, and we had the average team size based on the number of collaborators in all publications. Finally, we had the median journal prestige for each scientist across their whole career, showing the different lifetime publication trajectories (see Kwiek & Roszka, 2024).

We used a convenience sampling method, with an equal probability of selection method (Hibberts et al., 2012, p. 55). This means that every scientist who appeared in Scopus-indexed publications had the same chance of being included in the sample. We sent personalized invitations to all 65,300 researchers for whom Scopus provided an email address in early 2023. This group formed the survey population. We did not systematically exclude any groups of scientists from the sampling frame (Bryman, 2012, p. 187). At the same time, we cannot determine exactly how respondents differ from nonrespondents. For this reason, we cannot precisely assess the scale of potential nonresponse bias (Stoop, 2012, p. 122).

As in any survey, the key issue is sample representativeness. In this study, we assume that coverage error is relatively small. The majority of our sample includes scientists from disciplines in which publishing journal articles (indexed in Scopus) is the dominant form of scientific communication. This applies especially to engineering and technology (ENGTECH), natural sciences (NATSCI), and medicine (MED). Scientists from these three fields comprise the majority of the sample.

A substantial part of our sample (14%) comes from social science (SOC) scholars. This allows us to analyze perceptions of academic success beyond strictly STEM fields. Humanities (HUM) are less represented in the sample (9.3%). This reflects publication practices in this field. In the Polish humanities, Scopus-indexed journals play a smaller role.

The aim of the study is not to compare the response levels of disciplinary groups. Our aim is to identify links between different aspects of academic success. We also examine how these aspects interact with key demographic and professional variables and respondents' subjective assessments of success. In this context, possible underrepresentation of some disciplines among the research-active population of Polish scientists should not systematically distort the structure of relationships.

## 3.3. Data and study sample

Scientific disciplines were provided by respondents directly in the survey. In the questionnaire, respondents selected one of ten fields that follow the current Polish discipline classification (engineering and technology, medical sciences, health sciences, agricultural sciences, humanities, natural sciences, social sciences, veterinary sciences, theological sciences, and the arts).

We clustered original disciplines into five broader groups. These are ENGTECH (engineering and technology), MED (medical and health sciences), NATSCI (natural sciences, including agricultural and veterinary sciences), HUM (humanities, theological sciences, and the arts), and SOC (social sciences). This aggregation ensured sufficient numbers of observations in each category. It also ensured the comparability of the main areas of the science system. The final sample structure by aggregated disciplines is presented in Table 1.

**Table 1.** Distribution of basic sample characteristics

| Block | Category | Total | Total_colpct | Men | Women |
|---|---|---|---|---|---|
| Age group | Total | N=10848 | % (col) | N=5887 | N=4961 |
| | ≤29 years | 186 | 1.7 | 102 | 84 |
| | 30–39 years | 2576 | 23.7 | 1315 | 1261 |
| | 40–49 years | 3599 | 33.2 | 1786 | 1813 |
| | 50–59 years | 2438 | 22.5 | 1268 | 1170 |
| | 60–69 years | 1370 | 12.6 | 877 | 493 |
| | ≥70 years | 679 | 6.3 | 539 | 140 |
| Academic position | Total | N=10848 | | N=5887 | N=4961 |
| | Assistant Professor | 4834 | 44.6 | 2319 | 2515 |
| | Associate Professor | 3147 | 29.0 | 1763 | 1384 |
| | Full Professor | 1992 | 18.4 | 1382 | 610 |
| | Research Assistant | 875 | 8.1 | 423 | 452 |
| Instit ution | Total | N=10566 | | N=5734 | N=4832 |
| | IDUB University | 3284 | 31.1 | 1941 | 1343 |

| | Other Institutions | 1875 | 17.7 | 1014 | 861 |
|---|---|---|---|---|---|
| | Non-IDUB University | 5407 | 51.2 | 2779 | 2628 |
| Discipline | Total | N=10848 | | N=5887 | N=4961 |
| | ENGTECH | 2790 | 25.7 | 1941 | 849 |
| | HUM | 1010 | 9.3 | 483 | 527 |
| | MED | 1794 | 16.5 | 726 | 1068 |
| | NATSCI | 3731 | 34.4 | 2044 | 1687 |
| | SOC | 1523 | 14.0 | 693 | 830 |
| Academic age | Total | N=10564 | | N=5733 | N=4831 |
| | 0–9 years | 3911 | 37.0 | 1877 | 2034 |
| | 10–19 years | 3925 | 37.2 | 2072 | 1853 |
| | 20–29 years | 1848 | 17.5 | 1109 | 739 |
| | ≥30 years | 880 | 8.3 | 675 | 205 |

# 4. Results

## 4.1. Analysis of interrelationships among success dimensions

Figure 1 shows the polychoric correlation matrix for the ten dimensions of academic success ordinal responses measured on a 5-point Likert scale. Polychoric correlations allow us to treat the responses as observations of latent variables that are continuous. This approach is especially useful for subjective ratings. It is also better than rank correlations when we want to estimate the strength of relationships between unobserved constructs behind survey responses.

The heatmap shows that most correlations are positive but moderate ($\rho \approx 0.20$–$0.60$) and there are no very strong correlations ($\rho > 0.70$). This pattern suggests that success dimensions are related (and partly complementary) and cannot be reduced to a single dimension. Several clusters are clearly visible. The most strongly connected cluster is related to international scientific visibility. That cluster includes publications in top international journals, citations, international contacts, and invitations to keynote talks. A second cluster includes elements of institutional recognition and formal status. Publications in top national journals are also distinct. They are more weakly connected to internationalization-related dimensions. Even at this stage, academic success has a complex structure.

Based on the results of polychoric correlation matrix, we conducted an EFA (see Figure 2). The figure shows only relationships with an absolute value of at least $|0.3|$. We estimated the model using maximum likelihood, and we used an oblique oblimin rotation so that the factors could correlate. This rotation method often gives more realistic and better-fitting results in the social sciences than orthogonal methods. We determined the number of factors with parallel analysis. The result was clear: a four-factor solution is best; it is consistent and easy to interpret. The factor structure organizes the relationships visible in the correlation matrix. We interpret the first factor as academic prestige and visibility. It is strongly defined by membership in academies and associations, keynote talks, and affiliation with prestigious institutions ($\lambda \approx 0.64$–$0.67$). The second factor describes international publication productivity. It is dominated by publications in top international journals ($\lambda \approx 1.00$) and, to a lesser extent, citations ($\lambda \approx 0.31$). The third factor covers national recognition and research funding, and the fourth factor is mainly linked to formal career advancement. Publishing in top national journals has a negative loading on this factor ($\lambda \approx -0.51$).

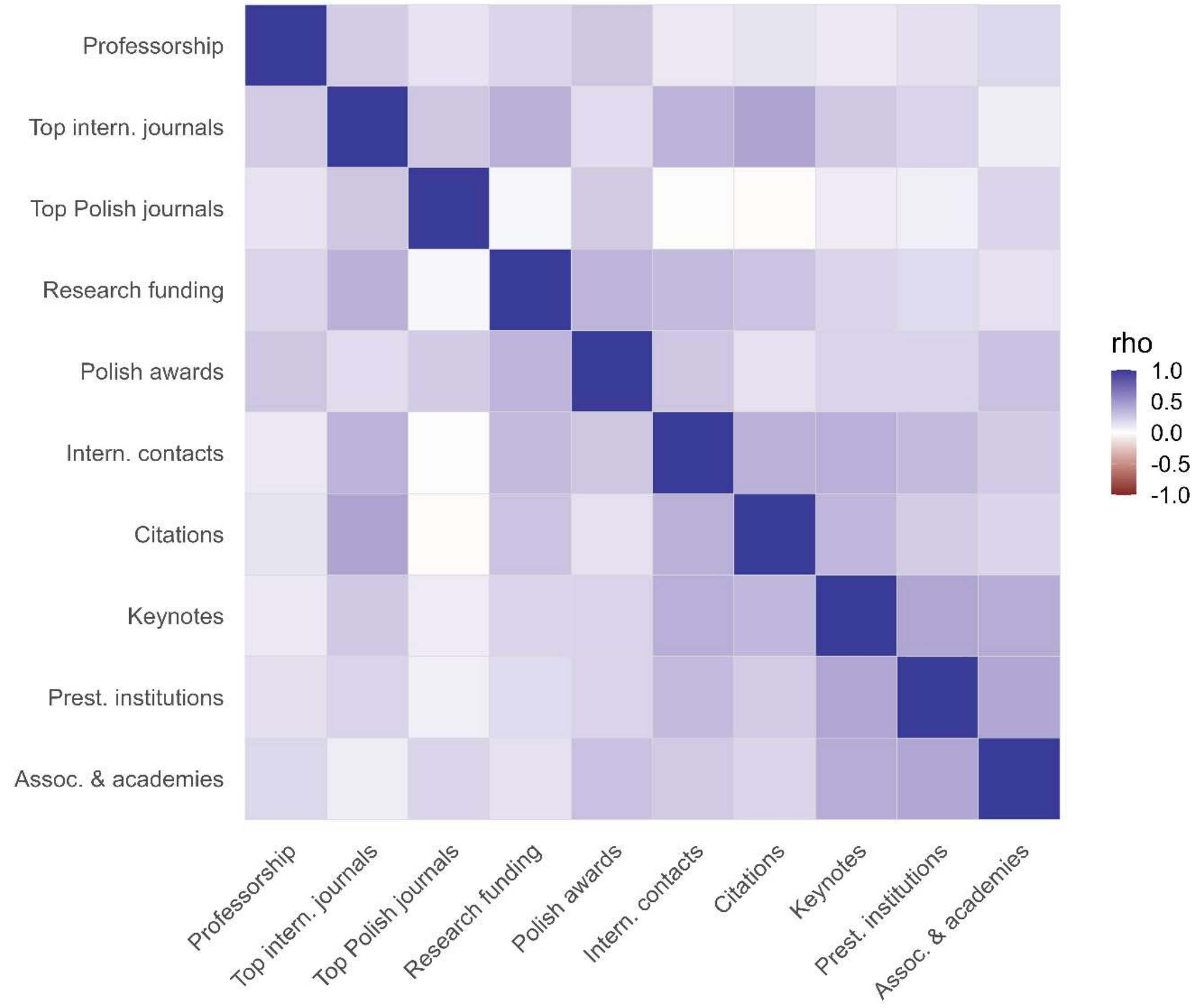

**Figure 1.** Polychoric correlations among success dimensions in a multidimensional perspective

Correlations between the factors are moderate (r ≈ 0.30–0.45). This confirms that the success dimensions are related. However, they do not reduce to a common latent construct.

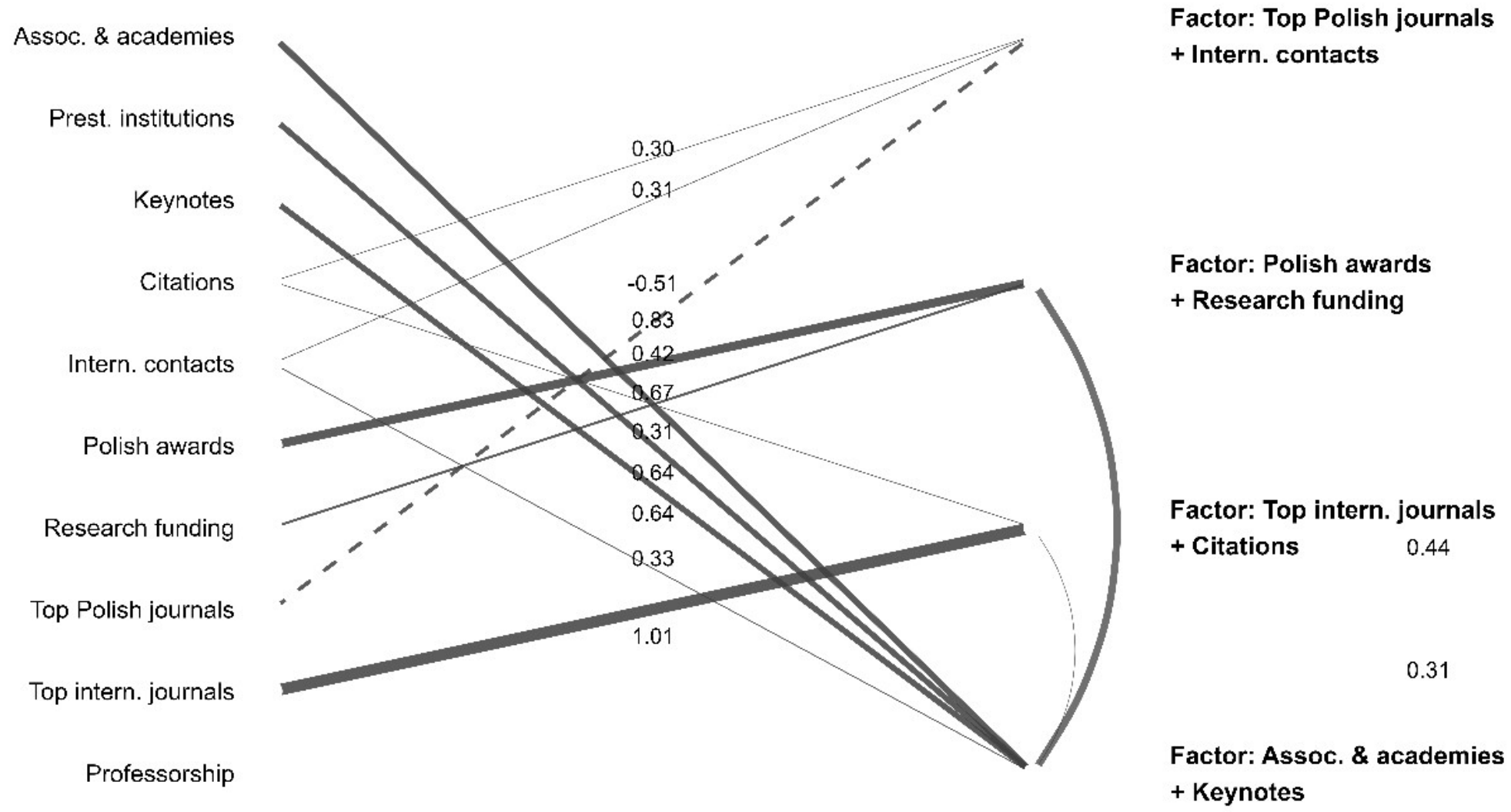

**Figure 2.** Results of factor analysis

Figure 3 shows strength centrality in the partial correlation network. In this network, links between pairs of dimensions are estimated while controlling for all other dimensions. Therefore, we can identify direct relationships and reduce the influence of indirect associations. Strength centrality is the sum of the strengths of a given dimension's direct connections with the rest of the system.

The highest centrality is observed for publishing in top international journals (strength ≈ 1.5). This indicates that it is a key hub that integrates different aspects of academic success. International contacts and membership in academies and associations also have high, but clearly lower, centrality (strength ≈ 1.1). These two dimensions link publication success with community and institutional recognition. The lowest centrality is observed for professorship (strength ≈ 0.6), which suggests that formal academic advancement is more weakly connected to other dimensions. It is more a career outcome than an element that binds the whole system.

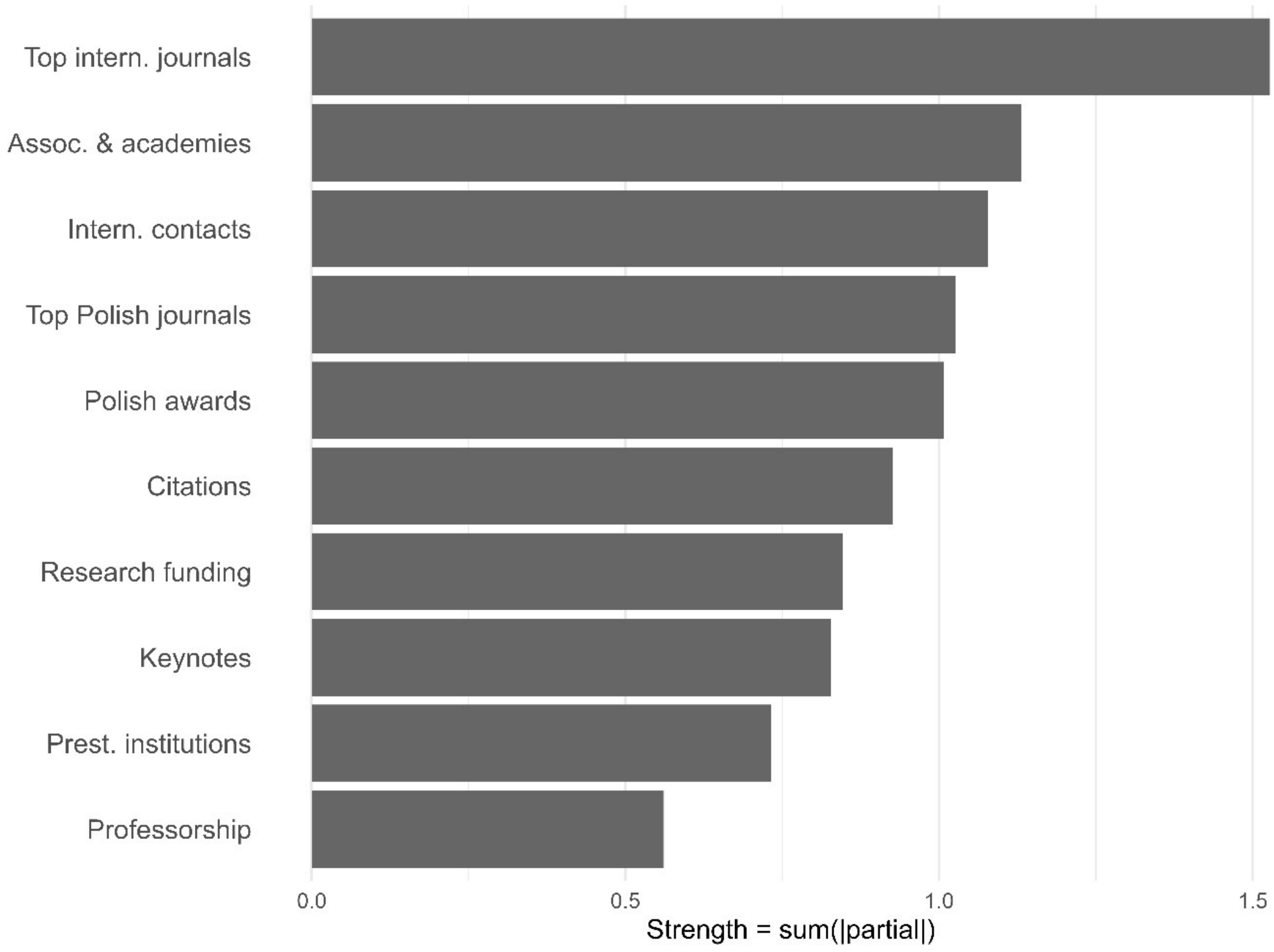

**Figure 3.** Partial correlations

The EBICglasso partial correlation network shows how the ten success dimensions are connected. EBICglasso allows us to build a network plot in which nodes represent variables and edges represent partial correlations between them. It removes spurious and weak links and keeps only the strongest relationships, which makes interpretation easier. It also helps avoid overfitting. We implemented EBICglasso using an R package. The node colors represent the four factors from the EFA. The highlighted node ("Top international journals") represents global publication prestige. It is the main hub in the network. It links international visibility, symbolic recognition, and the peripheral national cluster. This network structure is consistent with the earlier correlation and factor analyses, which also indicated that publication success in top international journals occupies a central position within the multidimensional structure of academic success.

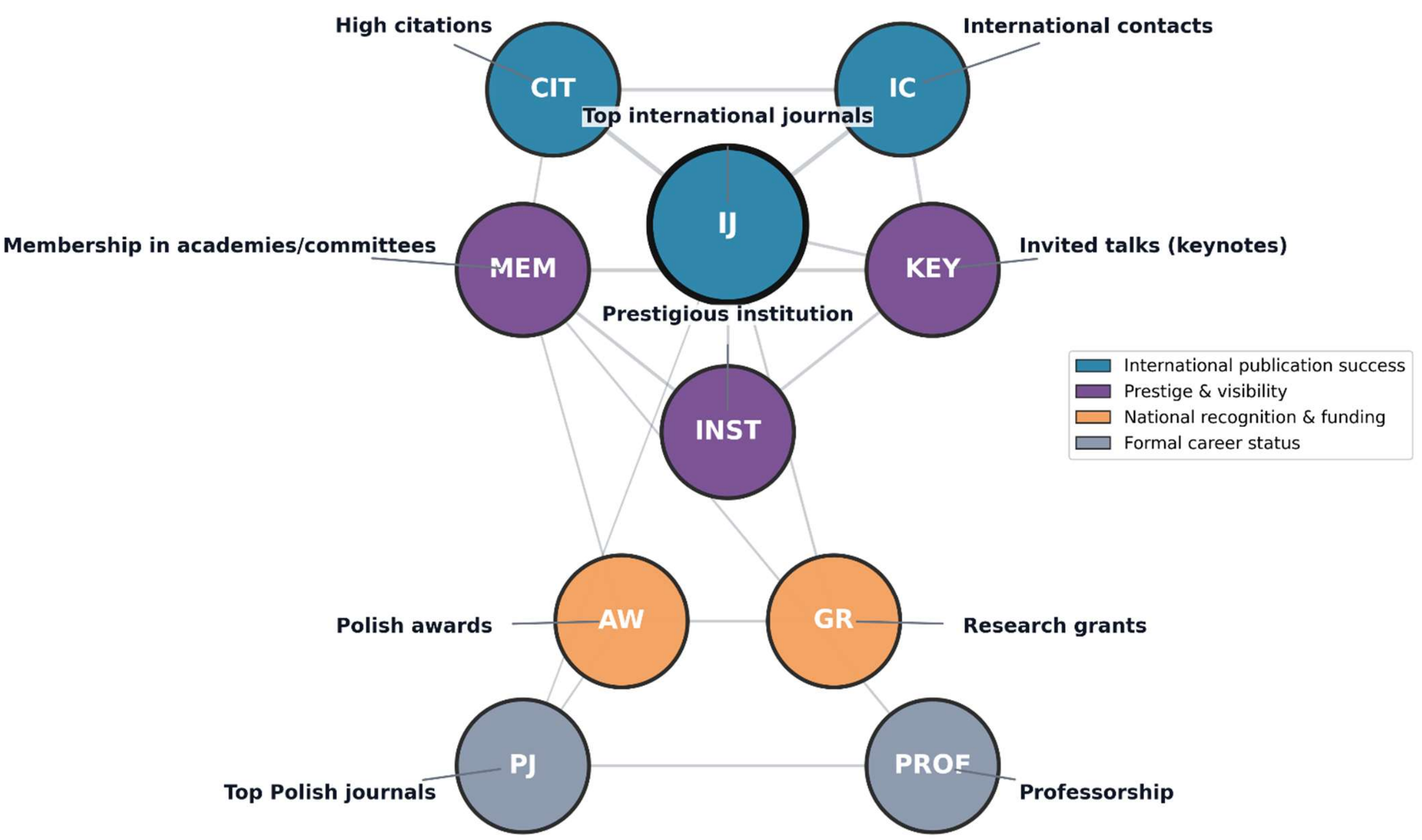

**Figure 4.** Partial correlation network of success dimensions (EBICglasso); node colors indicate EFA factor membership

Figure 5 shows the full network of relationships among the dimensions of academic success. We estimated this network using EBICglasso ($\gamma = 0.50$). This method combines partial correlations with regularization. Thus, it reduces the number of weak and potentially unstable connections. Nodes represent individual success dimensions, and edges represent direct relationships. Edge thickness reflects the strength of these relationships.

The network is dense, which means that there are many direct links among the analyzed dimensions. The strongest connections cluster around three elements: publications in top international journals, citations, and international contacts. These elements form the core of the network. However, many edges have moderate strength. This makes a simple, selective interpretation difficult. For this reason, in the next part of the analysis we use centrality measures and threshold-based approaches.

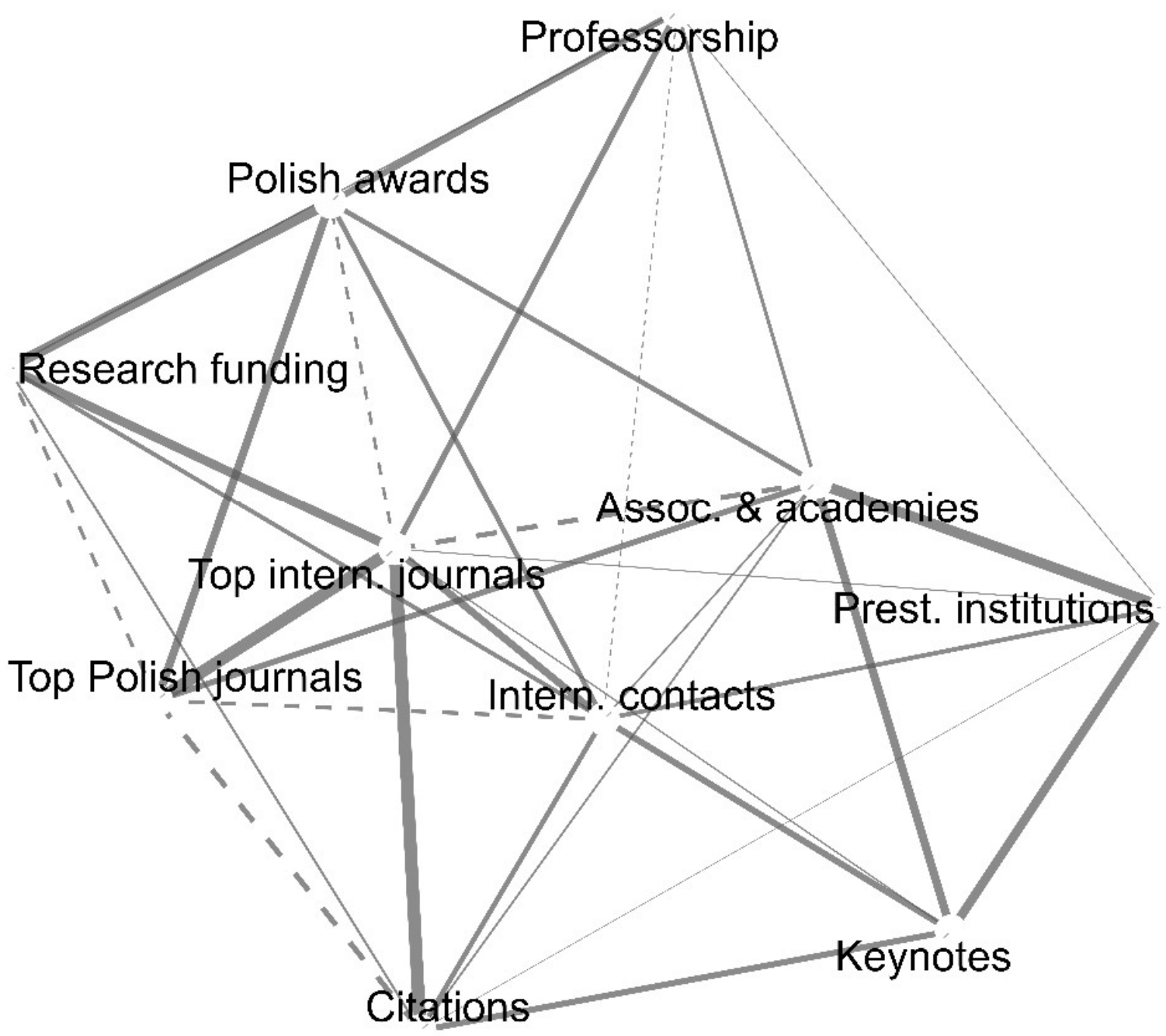

**Figure 5.** Partial correlation network (EBICglasso)

The MDS map (see Figure 6) was built from a distance matrix. MDS is a visualization tool, and it shows similarity or difference between objects in a two- or three-dimensional plot. Its goal is to simplify complex, multidimensional data by placing objects on a plane so that the most similar objects are close to each other, and the most different objects are far apart. We defined distance as $1 - |\rho|$, where $\rho$ is the polychoric correlation between success dimensions. This definition is easy to interpret. Strongly related dimensions are placed close together, and more weakly related dimensions are placed farther apart. The two-dimensional projection allows a quick assessment of similarities and differences between dimensions, while preserving the structure of pairwise distances as closely as possible.

The MDS map shows clear clustering of success dimensions, consistent with earlier analyses. Elements related to international scientific visibility are close to each other. Professorship and publications in top national journals, in contrast, are clearly farther away. Awards and research funding have an intermediate position. They connect international visibility with institutional (national) recognition. These results strengthen the claim that there are several different, though partly connected, logics of assessing academic success.

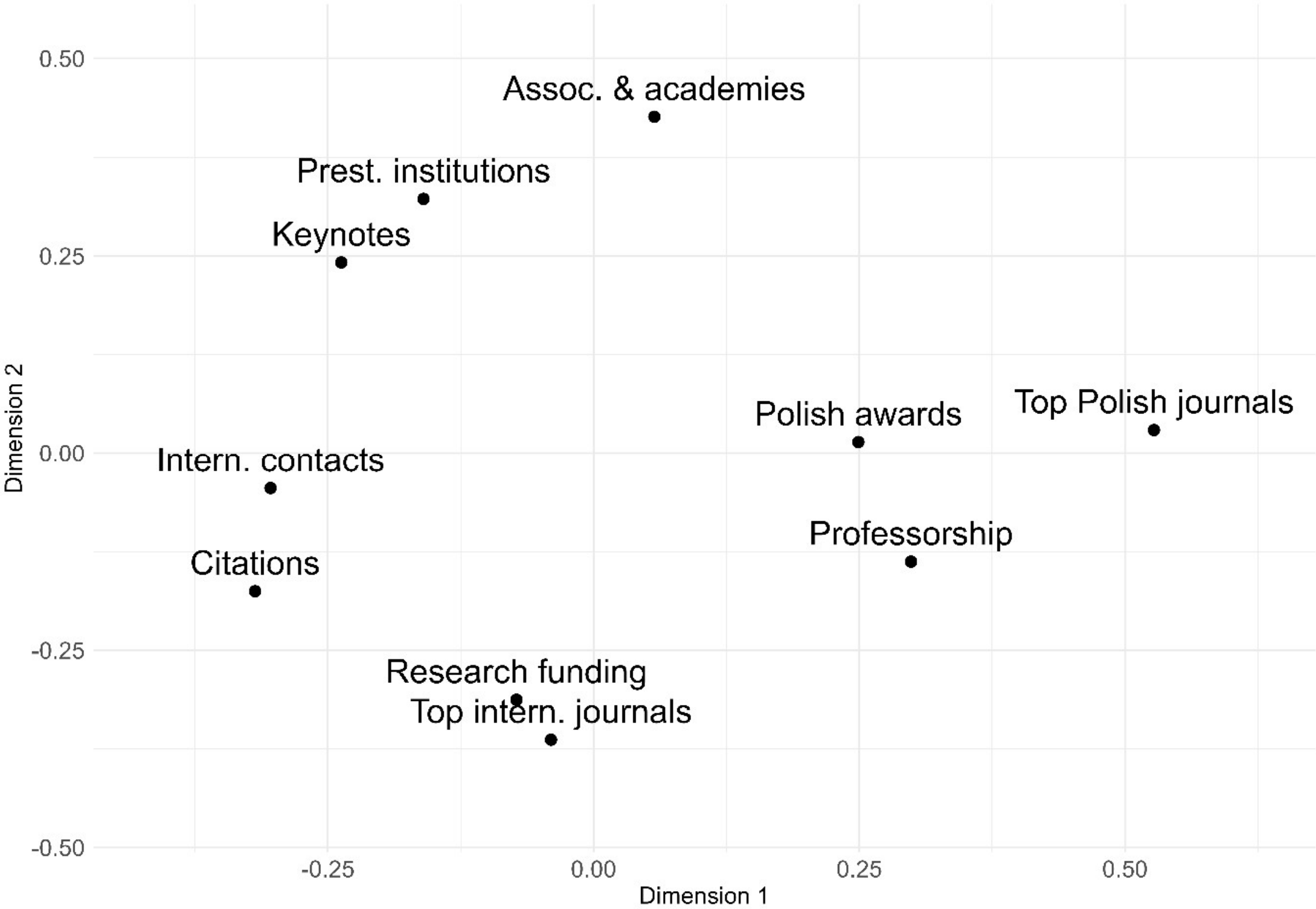

**Figure 6.** Multidimensional scaling map

Our analyses demonstrate that academic success is perceived as a multidimensional phenomenon. Polychoric correlations between the dimensions are mostly positive but moderate ($\rho \approx 0.20$–$0.60$). We do not observe very strong correlations ($\rho > 0.70$). This means that the different aspects of success are related. However, they are not substitutes for a single measure. Respondents view success as co-occurring but distinct achievements.

The most coherent area is international scientific visibility. Publications in top international journals, the number of citations, international contacts, and keynote invitations have the highest mutual correlations ($\rho \approx 0.45$–$0.60$), forming a clear separate cluster. Factor analysis confirms that this dimension is distinct. It identifies a factor dominated by publications in top international journals ($\lambda \approx 1.00$) and, to a lesser extent, citations ($\lambda \approx 0.31$). Global publication visibility emerges as a central element in definitions of scientific success.

In parallel, a dimension of prestige and community recognition emerges. It includes membership in academies and associations, affiliation with prestigious institutions, and plenary talks. These variables have high factor loadings ($\lambda \approx 0.64$–$0.67$). This indicates their shared reputational and symbolic meaning. Correlations between this factor and international productivity are moderate ($r \approx 0.30$–$0.45$), which means that community prestige and publication visibility are related but not the same.

Partial correlation analyses identify key elements of the direct dependence structure. The highest centrality is observed for publishing in top international journals (strength $\approx 1.5$). It clearly exceeds the other dimensions, which means that this aspect is the most strongly embedded in the whole system. It is directly linked to many other forms of success. International contacts and membership in academies and associations also have high, but lower, centrality (strength $\approx 1.1$). These elements act as intermediaries connecting publication visibility with institutional recognition.

Success dimensions linked to national recognition and research funding have an intermediate position. National awards ($\lambda \approx 0.83$) and research funding ($\lambda \approx 0.42$) are positively linked to both international visibility and institutional prestige. However, they do not form the core of direct network dependencies, which suggests their more instrumental role. They support other forms of success rather than define success on their own.

At the other pole is professorship (as a national achievement). It has the lowest centrality (strength ≈ 0.6), which indicates that formal academic advancement is more weakly connected to other success dimensions. It reflects long-term accumulation of achievements rather than a current position in the network. Similarly, publications in top national journals show relatively weak links to internationalization ($\rho \approx 0.00$–$0.20$). This suggests a partly separate, local logic of evaluating achievements.

The MDS map highlights these observations. Dimensions related to international visibility are adjacent. Professorship and publications in top national journals, in contrast, are clearly farther away. This confirms that different, though partly connected, pathways to and criteria of academic success exist.

Academic success is multidimensional, internally connected, and hierarchically organized. Individual career aspects do not operate as independent criteria. They form a coherent system of meanings. In this system, some elements are central, while others are peripheral or instrumental.

At this stage, we ask what differentiates this system of assessments. This concerns individual, professional, and institutional characteristics. Therefore, in the next step we move to bivariate analyses. The goal is to identify basic, selective links between success dimensions and single predictors.

Importantly, the same structural pattern appears consistently across correlation, factor, network, and MDS analyses.

## 4.2. Bivariate analyses

The bivariate analyses were exploratory: the goal was to identify basic links between each dimension of academic success and the characteristics of single researchers. We assessed relationships using Spearman’s rank correlation, which allowed us to use a single consistent approach. The approach was nonparametric, and it did not depend on the measurement level of predictors.

For binary variables, Spearman’s coefficient does not mean a classic correlation between two continuous variables. It represents a standardized difference in ranks between two groups, allowing us to compare the effect sizes of predictors measured on different scales. It also allows us to treat $\rho$ as a consistent effect size measure. It is robust to non-normal distributions and to heteroscedasticity.

The results of this part of our research are descriptive. We treat them as a reference point, an introduction to further multivariate analyses. In those analyses, we can account for the co-occurrence of multiple predictors.

Relationships between single categorical predictors and success dimensions are selective. None of the analyzed factors differentiates all aspects of success at once. This confirms the multidimensional character of academic success.

The most consistent pattern concerns the language of publication. Publishing in English is positively and significantly related to international dimensions of success. This applies especially to publications in international top journals, the number of citations, and keynote invitations. Typically, we observe ρ in the range of 0.25–0.40 (after False Discovery Rate (FDR) correction). We do not see similar relationships for national success. This indicates a clear separation between prestige circuits.

Disciplinary differences also matter, although they are more heterogeneous. The humanities are more often linked to national channels of recognition and to institutional prestige. In contrast, medical and technical sciences are more strongly embedded in the logic of internationalization and citations. This suggests that definitions of success partly depend on disciplinary specifics.

Academic position mainly differentiates formal and symbolic aspects of success. This concerns, above all, professorship, affiliation with prestigious institutions, and membership in academies. These relationships are cumulative, which suggests that they result from a long-term accumulation of achievements. They are not a simple effect of current publication activity.

Taken together, these patterns suggest that different predictors are linked to different dimensions of success rather than to academic success as a single, unified outcome.

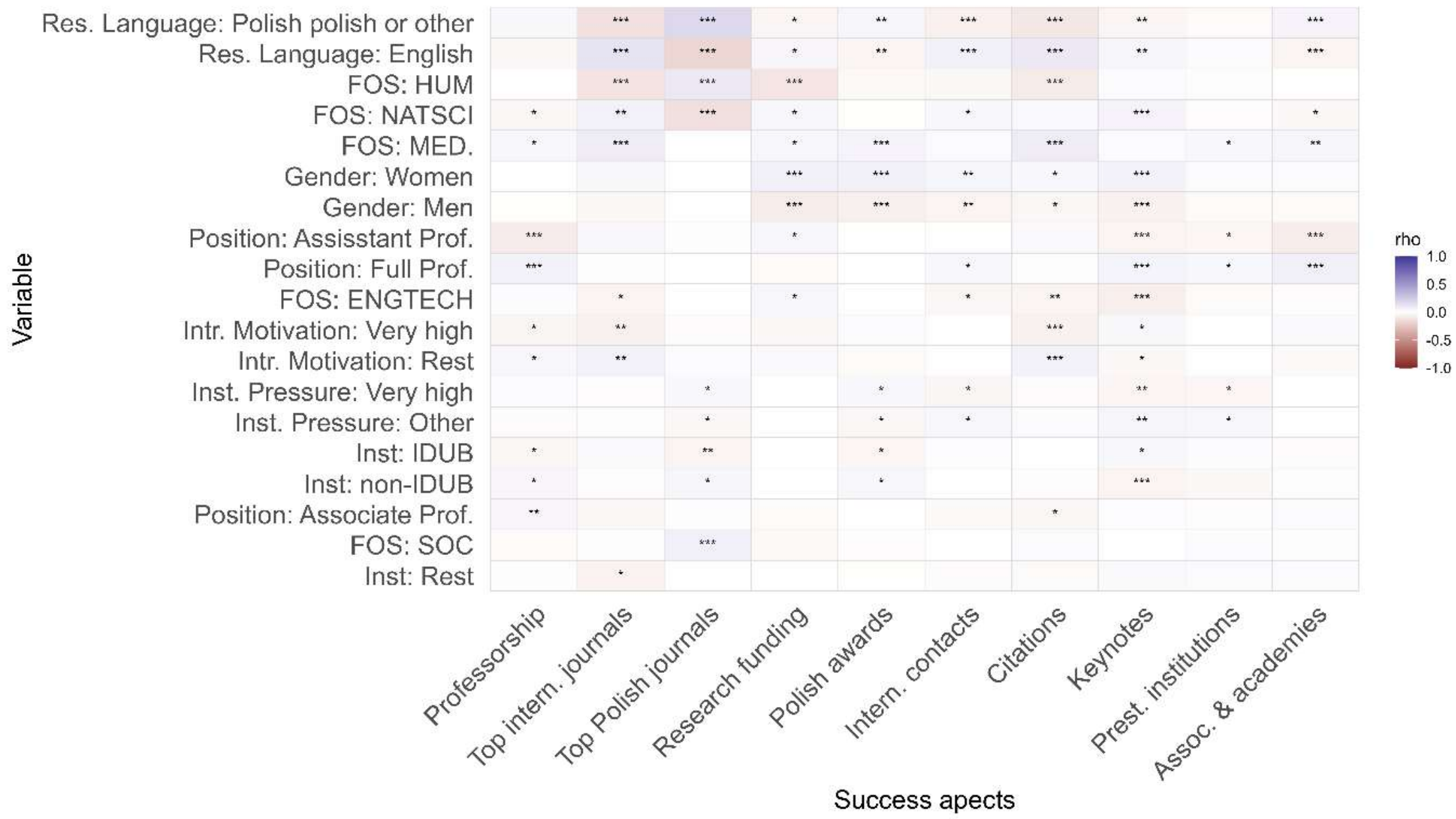

**Figure 7.** Bivariate relationships between success dimensions and single categorical predictors

Among numeric predictors, the strongest and most consistent relationships concern internationalization and the intensity of research activity. The international collaboration rate is positively correlated with most international dimensions of success. Spearman's coefficients are usually moderate (ρ ≈ 0.30–0.40).

The number of publications is significantly related to bibliometric visibility. This applies to publications in top journals and to citations. However, the links between publication counts and other forms of success are weak or not significant (e.g., institutional prestige and awards). This may suggest that high productivity matters in achieving success but is not sufficient to achieve broader recognition.

Biological age and academic age show a mixed pattern of relationships. Older age increases formal and cumulative achievements. This is visible, for example, for professorship and membership in academies ($\rho \approx 0.30$). However, older age is negatively correlated to internationalization and publishing in competitive channels. This points to cohort effects and suggests that career strategies change over a life course in science.

Overall, the pattern of relationships indicates that different mechanisms underlie different dimensions of academic success: international collaboration supports network-based visibility, publication output reflects productivity-based success, and age-related variables capture cumulative prestige accumulated over the course of a scientific career.

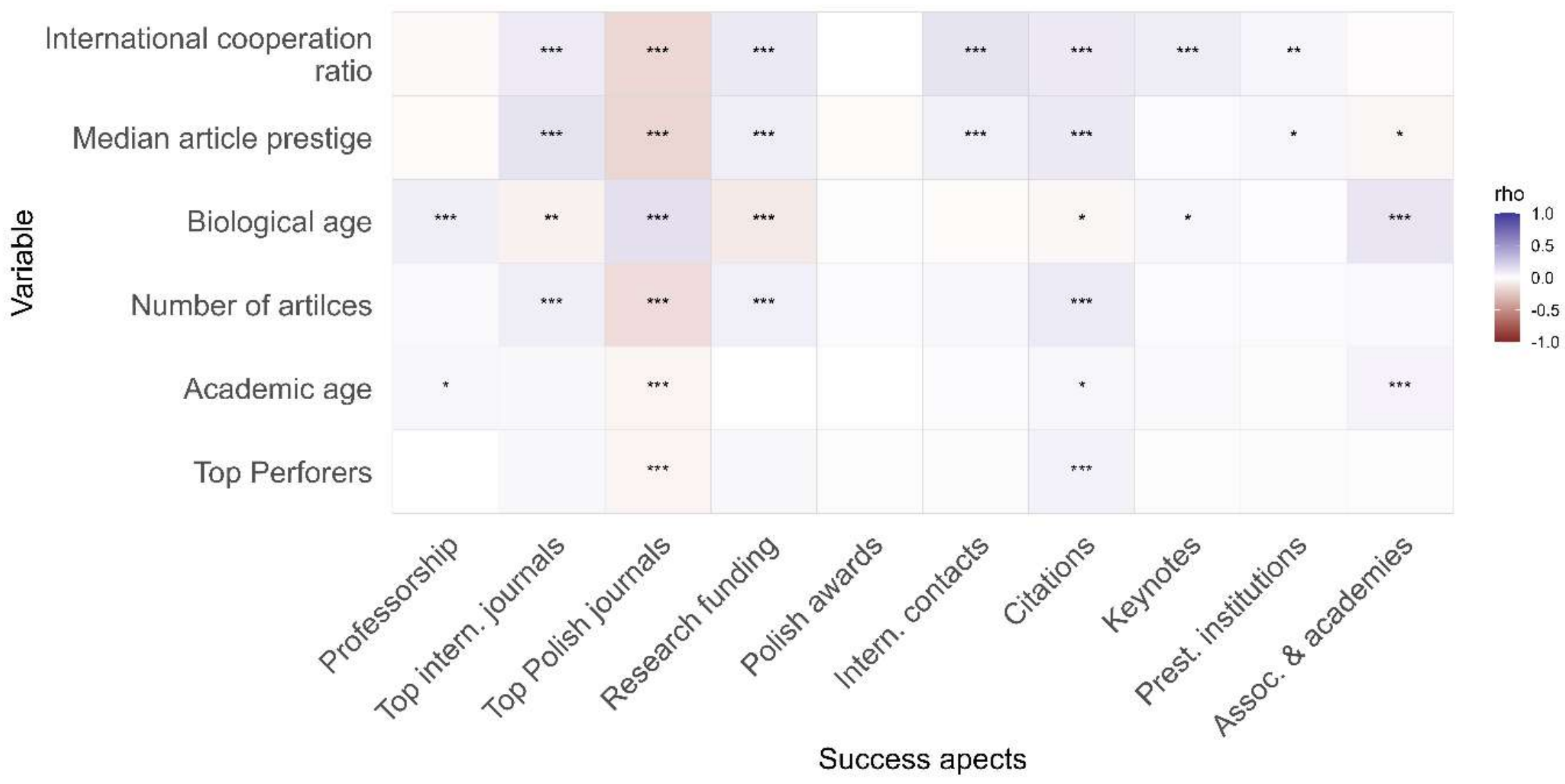

**Figure 8.** Relationships between success dimensions and single numeric predictors

Relationships between categorical predictors and factor scores (see Figure 9) form a clear pattern. Factors related to internationalization—publications in top international journals, citations, and international contacts—are significantly associated with the language of publication and with disciplinary affiliation. Factors capturing institutional prestige and community recognition, in contrast, show weaker relationships with these characteristics and are more selective.

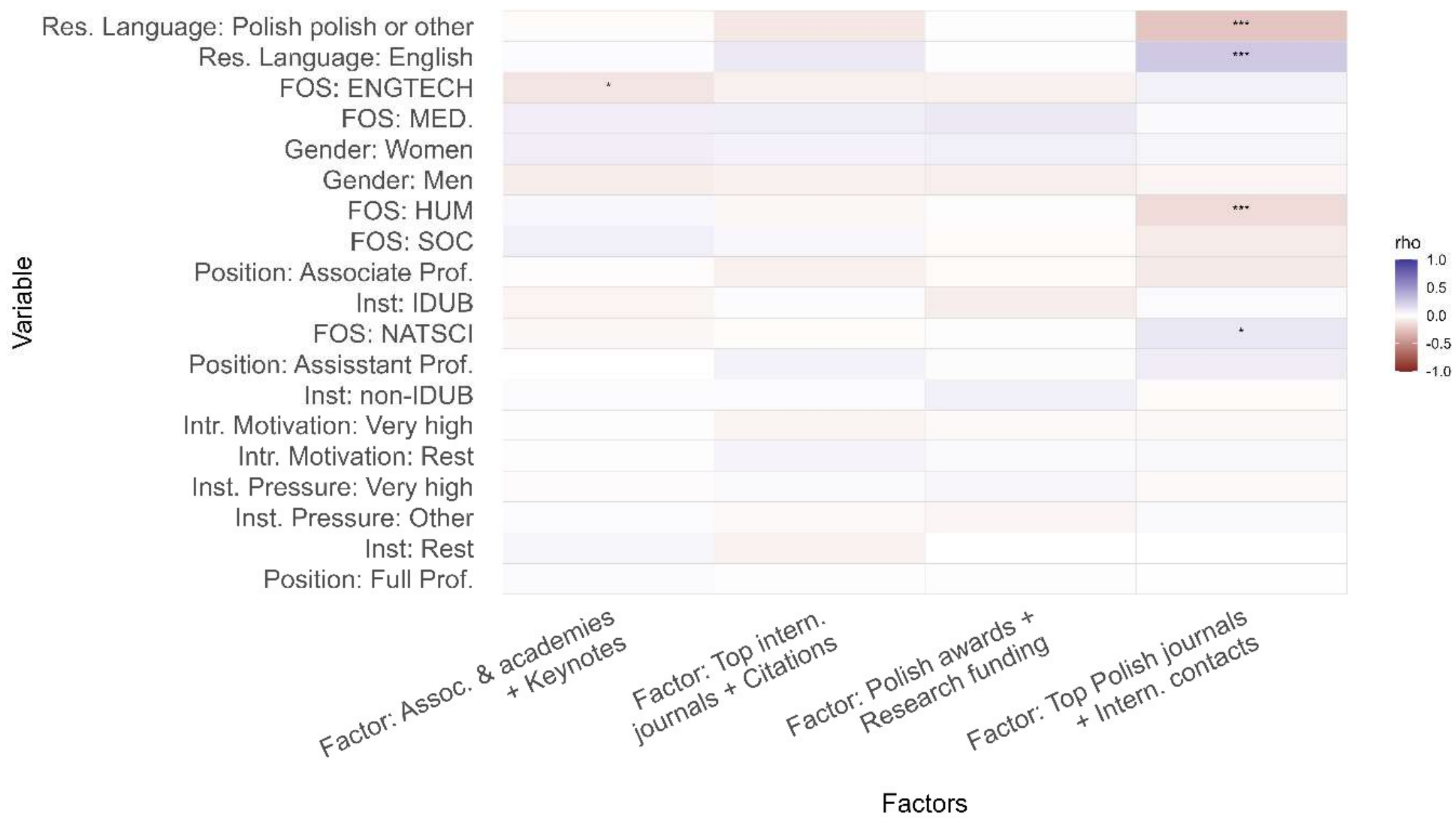

**Figure 9.** Relationships between single categorical predictors and factor scores from exploratory factor analysis

For numeric predictors (see Figure 10), the strongest and clearest relationships concern the factor of international publication visibility. This factor increases with the intensity of international collaboration. It also increases with bibliometric prestige and the size of a researcher's output. In contrast, factors related to awards, research funding, and institutional prestige are more weakly related to simple measures of research activity. Comparing results for categorical and numeric predictors shows one more point. The EFA factors differ not only in their structure but also in their sensitivity to the type of predictor.

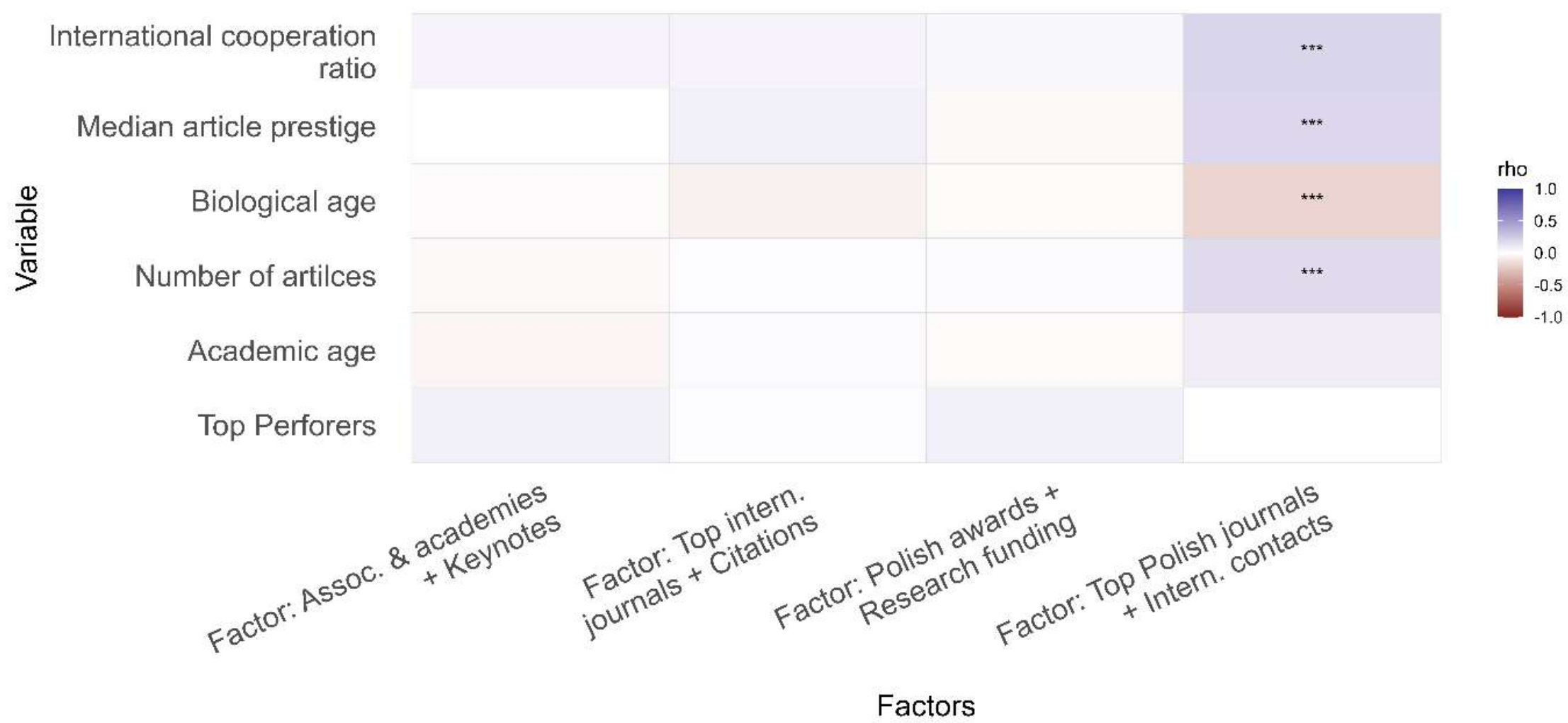

**Figure 10.** Links between quantitative predictors and factor scores from exploratory factor analysis

Bivariate analyses confirm that academic success is not uniform: it consists of several distinct dimensions only partly related. The strongest and most stable relationships concern internationalization and bibliometric visibility. For these measures, Spearman's coefficients are usually moderate (around 0.30–0.40).

Structural relationships also matter. This concerns discipline, academic position, and institutional affiliation. However, these relationships are selective. Their direction and strength differ across success dimensions. In particular, formal forms of recognition and institutional prestige do not follow directly from the scale of publication output.

Individual and motivational characteristics have weaker explanatory power in bivariate analyses: their role becomes visible only in more complex configurations. Therefore, these results justify moving to multivariate analyses that allow us to separate spurious from independent effects. They also allow us to better understand the mechanisms that produce different profiles of academic success.

Bivariate analyses also show that relationships are selective. No single predictor differentiates all success dimensions at the same time. Relationships vary in both direction and strength. This strengthens the conclusion that success is a configuration of several forms of recognition and not a single construct.

The bivariate approach has limitations as well. We cannot determine whether the observed relationships are independent and whether they result from the co-occurrence of other respondent characteristics. It does not fully account for the fact that ratings of different success dimensions come from the same individuals. Therefore, we move to a modeling approach that allows us to include multiple predictors simultaneously. It also allows us to model responses as part of coherent systems of meanings.

The results indicate that academic success is structured as a system of partly connected dimensions shaped by different mechanisms and predictors.

## 4.3. Model approach

Perceiving success in science is a complex, internally coherent phenomenon. Respondents do not evaluate each aspect of success in isolation from others. Their answers form a relatively coherent belief system that reflects professional experience, academic position, and individual views of what a successful scientific career is. Therefore, an approach based on many independent models (one model for each dimension) has limitations. It does not capture the relational and hierarchical nature of the phenomenon.

We use a modeling approach that addresses this limitation. We treat different aspects of success as repeated ratings given by the same person. This directly accounts for dependence among responses from one respondent. It also accounts for the fact that respondents differ in their general tendency to give higher or lower ratings. A random effect at the individual level captures this tendency. It works regardless of whether the item concerns publication visibility, research funding, or institutional position. This allows us to separate a respondent's overall "threshold of requirements" from the relative importance they assign to specific career elements.

We use GLMMs. GLMMs are an advanced class of statistical models. They combine features of generalized linear models and linear mixed models. GLMMs are needed when data are non-normal. This includes binary outcomes, count outcomes, or skewed outcomes. They are also needed when observations are correlated and therefore not independent. They are also useful when data are unbalanced, for example due to missing values or different numbers of observations per unit.

In the model, we keep the distinctiveness of the success dimensions. The model does not assume that all career elements have the same meaning. Instead, it allows us to show that some dimensions are rated as more central. Other dimensions are rated as secondary or peripheral. Importantly, this hierarchy is stable in the studied population.

The definition of the dependent variable is also important. The ordinal model, based on the 5-point scale, identifies factors that shift the entire distribution of ratings. It therefore shows what shapes the overall hierarchy of success evaluations. This is the most restrictive specification. It assumes that a given factor operates similarly at all scale levels. For this reason, structural factors are most visible in the model. These include gender, career stage, and institutional context. Such factors shape the overall tendency to give higher or lower success ratings.

However, perceptions of success are not always linear. For this reason, we add two binary models. We focus on the upper part of the distribution. One model targets high ratings. The second targets the highest ratings. This approach reveals mechanisms that may remain invisible in the ordinal model. Some factors do not shift the entire distribution. They start to matter only when we discuss very high success, especially elite success. This concerns, for example, research internationalization and publication prestige. In this sense, these factors operate selectively. They are not universal determinants of ratings.

Comparing the three specifications leads to a consistent conclusion. The mechanisms behind perceptions of success are threshold-based. What differentiates the overall hierarchy of ratings need not have to determine the assignment of the very highest rank. A lack of statistical significance in the ordinal model does not mean a lack of substantive importance. It may mean that a factor operates only at the extreme levels of the scale.

As a result, our approach allows us to analyze success as a multidimensional, hierarchical phenomenon. It also allows us to capture its internal differentiation. Instead of many separate models, we obtain one coherent picture. This picture shows which factors shape the general understanding of success. It also shows which factors operate selectively. These selective factors mark the boundary between a “good” and an “outstanding” scientific career. This interpretive coherence is the main value of the modeling approach.

**Table 2.** Parameters of the ordinal GLMM

| | OR | OR_low | OR_high | estimate | p.value |
|---|---|---|---|---|---|
| Academic age | 0.971 | 0.925 | 1.019 | -0.030 | 0.229 |
| International collaboration rate | 1.017 | 0.988 | 1.048 | 0.017 | 0.256 |
| Field: ENGTECH | 1.014 | 0.926 | 1.110 | 0.014 | 0.765 |
| Field: HUM | 0.935 | 0.833 | 1.048 | -0.068 | 0.248 |
| Field: MED | 1.242 | 1.121 | 1.377 | 0.217 | 0.000 |
| Field: NATSCI | 0.919 | 0.838 | 1.008 | -0.084 | 0.075 |
| Gender: Male | 0.601 | 0.568 | 0.636 | -0.509 | 0.000 |
| Institution: IDUB university | 0.977 | 0.863 | 1.107 | -0.023 | 0.719 |
| Institution: non-IDUB university | 1.010 | 0.895 | 1.141 | 0.010 | 0.866 |
| Very high institutional pressure (Q37_5=5) | 1.554 | 1.374 | 1.759 | 0.441 | 0.000 |
| Very high intrinsic motivation (Q37_4=5) | 1.243 | 1.176 | 1.314 | 0.218 | 0.000 |
| Total numer of articles | 1.045 | 1.005 | 1.088 | 0.044 | 0.028 |

| Success aspect: Elected membership in associations, academies, committees, etc. | 0.282 | 0.266 | 0.299 | -1.266 | 0.000 |
|---|---|---|---|---|---|
| Success aspect: Publications in top international scholarly journals | 6.527 | 6.116 | 6.965 | 1.876 | 0.000 |
| Success aspect: Publications in top Polish scholarly journals | 0.170 | 0.160 | 0.181 | -1.770 | 0.000 |
| Success aspect: Research funding / grants obtained | 1.268 | 1.196 | 1.345 | 0.238 | 0.000 |
| Success aspect: Polish distinctions and awards | 0.364 | 0.344 | 0.386 | -1.011 | 0.000 |
| Success aspect: Broad international contacts | 1.778 | 1.676 | 1.885 | 0.575 | 0.000 |
| Success aspect: High number of citations | 2.061 | 1.942 | 2.188 | 0.723 | 0.000 |
| Success aspect: Plenary talks at international conferences | 0.684 | 0.645 | 0.724 | -0.380 | 0.000 |
| Success aspect: Employment at a prestigious institution | 0.741 | 0.700 | 0.785 | -0.299 | 0.000 |
| Academic position: Assistant Professor | 0.725 | 0.652 | 0.807 | -0.322 | 0.000 |
| Academic position: Associate Professor | 0.817 | 0.745 | 0.897 | -0.202 | 0.000 |
| Median article prestige (lifetime) | 1.017 | 0.987 | 1.049 | 0.017 | 0.259 |
| Language of research: English | 0.969 | 0.910 | 1.031 | -0.032 | 0.317 |
| Top performers: Rest | 0.918 | 0.826 | 1.020 | -0.086 | 0.110 |
| Biological age | 1.067 | 1.020 | 1.116 | 0.065 | 0.004 |

The ordinal model (Table 2) shows how individual, professional, and institutional characteristics shape the overall tendency to give higher or lower ratings of academic success.   funding, and institutional position. We examine the questions: Who is more likely to treat different career elements as "high" or "very high" indicators of success? And who treats them more cautiously?

The strongest individual effect is gender. Being male clearly reduces the tendency to give high ratings. The odds of assigning a "high" or "very high" rating for men are about 40% lower than for women (OR = 0.601; 95% CI: 0.568–0.636; $p < 0.001$). This means that, with the same professional and institutional characteristics, men more often judge the analyzed career aspects as less "successful." This effect does not apply to one element only. It applies to the whole hierarchy of ratings. It indicates more restrictive success criteria.

A similar mechanism (albeit weaker) emerges for academic rank. Assistant professors are less likely to assign high ratings than full professors. Their odds of a "high" or "very high" rating are about 27% lower (OR = 0.725; 95% CI: 0.652–0.807; $p < 0.001$). Associate professors also rate more restrictively, but the effect is smaller. The difference is about 18% (OR = 0.817; 95% CI: 0.745–0.897; $p < 0.001$). As scientists move to more advanced career stages, expectations increase. What may look like "high" success early in a career may no longer meet the criteria for truly important success later.

We observe a different direction for age. A one-standard-deviation increase in age raises the tendency to give higher ratings by about 7% (OR = 1.067; 95% CI: 1.020–1.116; $p = 0.004$). Older respondents more often treat different career aspects as valuable, even after controlling for academic position. This can be interpreted as a change in perspective. Over time, success is judged less in a yes–no way. The criteria become broader and more inclusive.

Among variables related to institutions and motivation to conduct research, institutional pressure emerges as especially important. High institutional pressure increases the odds of giving high ratings. The odds of a "high" or "very high" rating increase by more than 55% (OR = 1.554; 95% CI: 1.374–1.759; $p < 0.001$). In highly demanding academic environments, success is perceived as more clearly defined, and career aspects are more often treated as key indicators of a "successful" scientific career.

Very high intrinsic motivation to pursue research works in a similar way. Scientists with such motivation more often give high ratings. The odds of a "high" or "very high" rating are about 24% higher (OR = 1.243; 95% CI: 1.176–1.314; $p < 0.001$). This effect applies to the entire hierarchy of ratings: publication visibility, funding, and institutional position. Motivation therefore strengthens the perceived importance of different career elements.

There is also a significant (but small) effect of publishing experience: a higher lifetime publication output is correlated with a slightly higher tendency to give high ratings in success assessments. A one-unit increase raises the odds of a "high" or "very high" rating by about 4.5% (OR = 1.045; 95% CI: 1.005–1.088; $p = 0.028$), which suggests that higher publishing intensity raises the reference point of scientists. Highly productive scientists more often treat career elements as important components of success compared to other scientists.

Medicine stands out clearly. Scientists in this field more often assign high success ratings. Their odds of a "high" or "very high" rating are about 24% higher than in the reference category (OR = 1.242; 95% CI: 1.121–1.377; $p < 0.001$). This means that, in medicine, the career aspects analyzed are more often seen as central to the definition of success. In other fields, the effects are weaker and often not statistically significant, suggesting more consistent rating patterns outside medicine.

It is also important to note the factors that do not shift the entire hierarchy of ratings in the ordinal model. This concerns, among others, the intensity of international collaboration, the prestige of journals, the language of research, and affiliation with a university included in an excellence program. These variables do not move the distribution of ratings toward higher or lower categories. This does not mean that they are substantively irrelevant. The results rather suggest that their effects are not linear. They become visible mainly at the highest rating levels. Binary analyses confirm this.

The model therefore shows that scientists treat success as a coherent hierarchy of ratings. Individual career aspects occupy relatively stable positions within it. Some elements are seen as central. Others are seen as secondary or peripheral. This hierarchy is shared by the whole population. Respondent characteristics do not change its structure. They change the overall level of ratings. In other words, they shape how strictly or inclusively success is defined.

The most important indicator of success is publishing in top international journals. This aspect has more than six times higher odds of receiving a high rating than formal academic advancement, which serves as the reference point (OR = 6.53; $p < 0.001$). The core of success also includes citations and international contacts. Both clearly increase the probability of high ratings (OR = 2.06 and OR = 1.78). They are treated as direct signals of impact, recognition, and presence in the international circulation of knowledge.

Grants matter, but their role is clearly secondary. They increase the odds of a high rating by about 27% (OR = 1.27). Their role is mainly instrumental. They are seen as a resource that helps achieve other goals. This concerns, above all, publishing and building visibility. Grants are therefore not treated as a stand-alone and sufficient marker of a successful career.

Lower in the hierarchy are employment at a prestigious institution and plenary talks at international conferences. Both have lower odds of a high rating than the reference point (OR = 0.74 and OR =

0.68). They are treated as accompanying elements of a career. They can strengthen image and visibility, but they are not seen as the foundation of success.

Even weaker importance is assigned to national awards and elected membership in academies and committees. Both clearly reduce the odds of a high rating (OR = 0.36 and OR = 0.28). The lowest position is held by publications in top Polish journals. They have the smallest odds of being treated as a central indicator of success (OR = 0.17). This points to the marginal role of national publication excellence as a stand-alone measure of academic success.

Formal academic advancement (full professorship) remains symbolically important. However, it is not a central criterion of success. Overall, the results show a shift in the definition of success toward international visibility, impact, and recognition. At the same time, the importance of formal, institutional, and national signals declines. Success is therefore strongly oriented toward the global circulation of knowledge and not local or formal confirmation of position.

The model also shows that perceptions of success are of a systemic nature. Respondents do not rate career dimensions in isolation: they relate them to a general orientation of what “true” academic success is. Factors such as gender, acavdemic rank, institutional pressure, and intrinsic motivation to perform research do not only affect single ratings. They shift the entire hierarchy of meanings. This distinguishes our approach from using a set of independent models for separate dimensions of success. It also captures success as a multidimensional but internally ordered phenomenon.

In contrast to descriptive and bivariate analyses, the modeling approach focuses on mechanisms that shift the entire hierarchy of ratings. It does not estimate separate models for individual career aspects but analyzes all success dimensions jointly within a single hierarchical framework.

### 4.4. Threshold structure and latent success dimension

The thresholds estimated in the ordinal cumulative logit model show where the boundaries between consecutive categories of the 5-point Likert scale lie. These boundaries refer to an underlying, continuous dimension of perceived success. The thresholds are ordered from lower to higher (from 1|2 to 4|5). This confirms that the model specification is correct and that the ordinal scale is appropriate.

**Table 2.** Estimated Likert-scale thresholds in the mixed ordinal (cumulative logit) model

| term | estimate | std.error | statistic | p.value |
|---|---|---|---|---|
| 1\|2 | -3.746 | 0.102 | -36.860 | 0.000 |
| 2\|3 | -2.520 | 0.101 | -24.956 | 0.000 |
| 3\|4 | -1.064 | 0.101 | -10.581 | 0.000 |
| 4\|5 | 0.566 | 0.101 | 5.632 | 0.000 |

Very low, strongly negative threshold values for the 1|2 (−3.75) and 2|3 (−2.52) transitions mean that the lowest responses were rare. Even at an average level of the latent “inclination,” respondents often chose at least “2” or “3.” In other words, the analyzed success dimensions were usually rated at least moderately high. The 3|4 threshold (−1.06) marks the boundary between neutral and high ratings. This is the range in which responses start to differentiate clearly. At this point, the explanatory variables matter most, shifting the probability toward higher ratings.

In contrast, the positive value of the 4|5 threshold (0.57) shows that the “5” response appears only at a high level of latent inclination to treat something as success. The highest category is therefore elite. It

also reflects a more restrictive criterion for recognizing an achievement as a fully realized dimension of success.

The statistical significance of thresholds is not the key point of interpretation. More important are their order, their signs, and the distances between them. These features show an asymmetric scale structure. Moving to the highest category is clearly more difficult than moving between two lower categories.

As a result, although many variables increase the probability of higher ratings, reaching category "5" requires a stronger shift in the latent success dimension. This justifies using additional, binary definitions of success in robustness analyses.

The forest plot (Figure 11) compares odds ratios with 95% confidence intervals from two binary GLMMs: "high success" (≥4) and "elite success" (=5). Differences between the models show a threshold mechanism. Institutional pressure and intrinsic motivation operate much more strongly for elite success. At the same time, the hierarchy of success dimensions is similar in both models. Global publication prestige remains dominant in both thresholds. The position of national publication channels remains peripheral regardless of the threshold.

These results indicate that perceptions of academic success are structured by a threshold mechanism in which different factors become influential at different levels of the success scale.

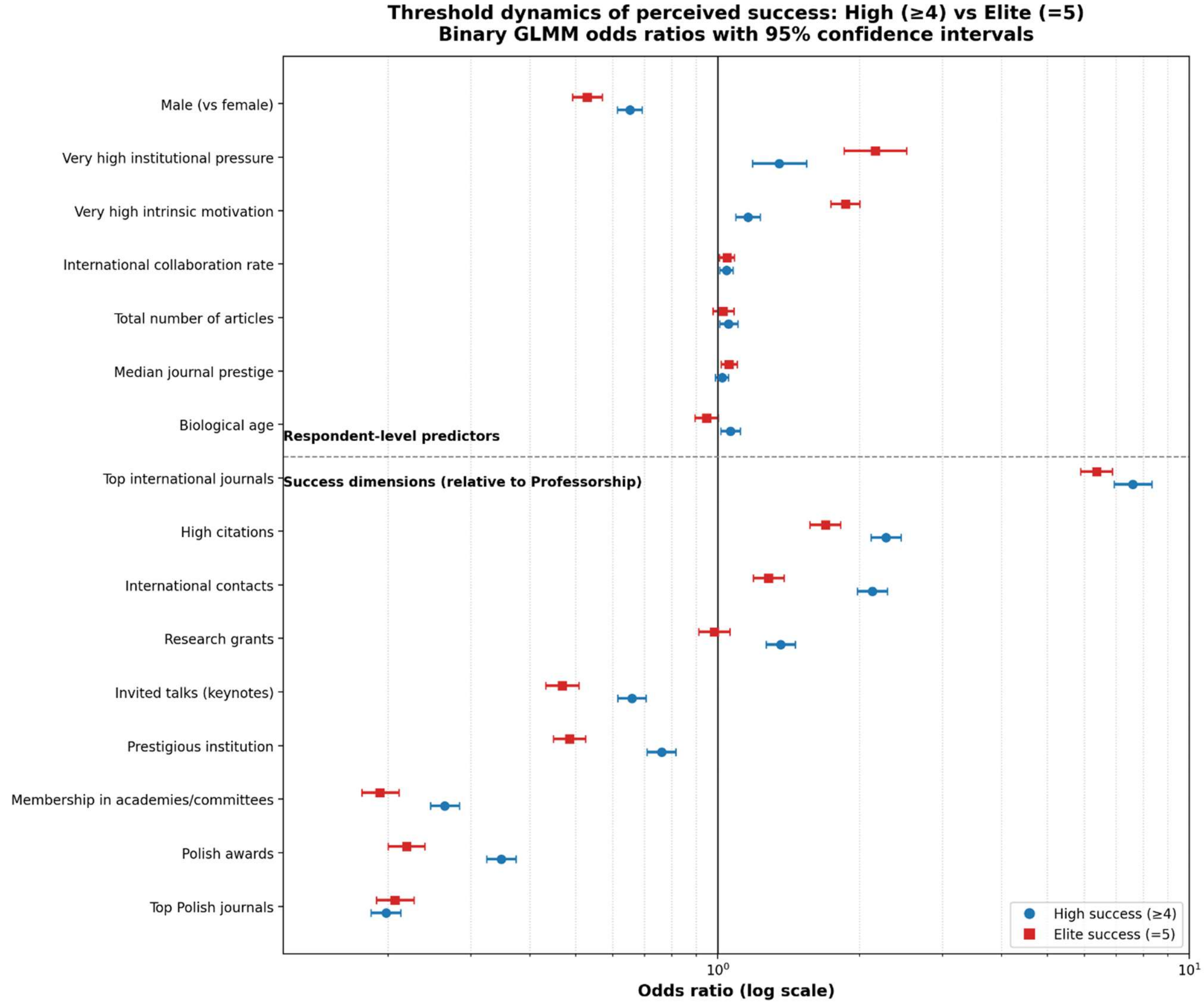

**Figure 11.** Threshold dynamics of perceived success: odds ratios (95% CI) comparing high (≥4) vs. elite (=5) success definitions in binary GLMMs

## 4.5. Robustness analyses: binary definitions of success

To test the robustness of the results, we used two binary definitions of success. In the first, the dependent variable meant a high rating (≥4), and in the second, it meant only the highest rating (=5). We estimated both models as mixed logistic models. We included a respondent-level random intercept, which allowed us to account for relationships across multiple ratings given by the same scientist (see Appendix, Tab1A).

Both models fit the data well. For the ≥4 threshold, the log-likelihood was −46,300 and AIC = 92,658, and for the =5 threshold, the log-likelihood was −40,085 and AIC = 80,227. The number of observations was the same in both models (n = 81,734).

Variance-based fit measures show that both fixed and random effects matter. In the ≥4 model, marginal $R^2$ was 0.21, and conditional $R^2$ was 0.42. The Interclass Correction Coefficient (ICC) was 0.26. This means that about one-quarter of the total variation in responses comes from differences between respondents. In the =5 model, heterogeneity was even stronger. Marginal $R^2$ was 0.22, conditional $R^2$ was 0.49, and ICC was 0.35, which suggests that maximum-success ratings vary especially strongly at the individual level.

At the level of fixed effects, both binary models lead to conclusions that are directionally consistent with the main model. Age was linked to a slightly higher probability of a high rating (OR ≈ 1.06 for ≥4). Men indicated high or maximum ratings less often than women (OR ≈ 0.65 for ≥4 and OR ≈ 0.53 for =5). The strongest and most stable effects concerned motivational variables. Very high intrinsic motivation almost doubled the odds of the highest rating (OR ≈ 1.86). Very high institutional pressure increased these odds by more than twofold (OR ≈ 2.16). The consistency of signs and the relative sizes of effects across the two specifications indicate that the main conclusions do not depend on the definition of high success selected.

### 4.6. Quality and adequacy of the modeling approach

The modeling approach has good fit and produces consistent results with different specifications. Binary models used in the robustness analyses achieve moderate marginal $R^2$ values (about 0.21–0.22), which means that fixed effects explain about one-fifth of the total variation in success ratings. At the same time, conditional $R^2$ increases to 0.42–0.49 after adding the respondent-level random intercept. High ICC values (0.26–0.35) confirm strong differences between respondents, which supports the use of mixed models. Simpler approaches would ignore intrapersonal dependencies.

The ordinal model, which is the main specification, also shows stable fit. The number of observations is large (n = 81,734). The log-likelihood was −109,502 (AIC = 219,068 and BIC = 219,366). In cumulative logit models, there is no single standard $R^2$ measure. Therefore, information criteria are crucial. They indicate a reasonable trade-off between model fit and model complexity. The model includes many predictors and a respondent-level random effect. There are no signs of convergence problems. Hessian and variance–covariance matrices are complete, which confirms the numerical stability of estimation.

An additional confirmation of correct model specification comes from the Likert-scale threshold pattern. The thresholds are ordered from lower to higher, from −3.75 (transition 1|2) to 0.57 (transition 4|5). There are no violations of monotonicity. This means that the 5-point scale works properly in this context. The positive value of the highest threshold also indicates that choosing category “5” requires a strong shift on the latent dimension of perceived success, which is consistent with the binary-model results. It also strengthens the interpretation of “5” as a restrictive definition of success.

To sum up, our results are consistent across the ordinal model and binary models. The directions and relative sizes of effects are aligned, and estimation is stable. The threshold structure is clear, which suggests that the conclusions are robust to the operationalization of the dependent variable rather than to an artifact of a particular model specification.

## 5. Discussion

Academic success emerges from this research not as a set of individual preferences but rather as an institutionalized structure of knowledge production. In this structure, different forms of recognition are relationally linked and hierarchically ordered (Cole & Cole, 1973; Whitley, 2000; Whitley & Gläser, 2007). In this sense, our study reveals the internal architecture of academic production. This architecture is multidimensional, but it is not symmetric. Not all forms of symbolic capital have the same weight or comparable convertibility (Bourdieu, 1988, 1993; Lamont, 2009). We show that success metrics do not operate as independent indicators but rather form a configuration of relationships. In their configuration, one dimension can act as a hub that integrates the whole system (Moed, 2005; Sugimoto & Larivière, 2018; Waltman, 2016).

Our empirical results demonstrate this in the factor structure and the network relations. Both approaches show that global publication visibility has the highest centrality. We operationalize this dimension mainly as publishing in top international journals. This dimension is associated with other areas of recognition such as citations, invitations to plenary talks, or committee roles (Borgatti et al., 2018; Bornmann & Daniel, 2008; Newman, 2001). Our conclusions are aligned with classic research on stratification and elites in science. In this line of research, the distribution of recognition is steep, and advantages are reproduced through feedback loops connecting visibility, access to resources, and further publication opportunities (Allison & Stewart, 1974; DiPrete & Eirich, 2006; Zuckerman, 1977). If global publication prestige operates as a core hub, then success in this dimension increases the chances of success in other dimensions. The point is not only additive accumulation, such as "more publications → more citations." The point indicates a relational centrality effect in a prestige network. Success in the core dimension then facilitates the flow of symbolic capital into other dimensions of recognition (Borgatti et al., 2018; Epskamp et al., 2018).

Our analyses suggest that, in the Polish context, symbolic international capital has higher convertibility than symbolic local capital. This is because Poland is strongly connected to the global circulation of scientific communication (to use Bourdieu's terms; Bourdieu 1983, 1988). Publications in highly selective international journals are therefore not only a measure of productivity: they also carry global recognition. And this recognition acts like a currency with high purchasing power in many segments of the field (Aksnes et al., 2019; Garfield, 2006; Sugimoto & Larivière, 2018). At the same time, local achievements look different (e.g., national publications or national recognition). They form a peripheral cluster among the dimensions of success, with weaker structural connections.

The literature on center–periphery relations in science emphasizes that the global circulation of rewards is unevenly organized (Kwiek, 2021; Kwiek et al. 2024). It also emphasizes that peripheral systems often need to internalize the center's criteria to gain visibility (Marginson, 2022; Schott, 1998). Our results support this claim at the micro-level of individual scientists. In their ratings, Polish respondents treat global publication prestige as the dominant dimension of success. The national field of knowledge production is powerfully influenced by global logics (Hazelkorn, 2015; Marginson, 2022).

Our results can be interpreted in terms of the "prestige economy" in science and the sociology of evaluation. Under intensified parameterization and inter-institutional comparison based on productivity and citation metrics, prestige becomes a scarce good. It is distributed through highly selective channels (Dahler-Larsen, 2012; Espeland & Sauder, 2007; Hazelkorn, 2015; Kwiek, 2021). Rankings and indicators do not only describe the academic reality—they also powerfully shape it. They generate strategic adaptations of institutions and individual scientists (Espeland & Sauder, 2007; Power, 1997; Strathern, 2000). In this context, the threshold pattern we observe is especially important. The highest success category (5 – elite success) requires crossing a qualitative threshold of recognition. This is consistent with research on elite stability and scientific "stardom" (Abramo et al., 2017; Kwiek & Szymula, 2025a; Kwiek & Roszka , 2025; Zuckerman, 1977). Threshold dynamics mean that differences at the top of the hierarchy are not a simple continuation of differences in the middle. The line of achievements is not linear, and reaching the top requires that a different logics be followed (e.g., *Nature* and *Science* among journals or European Research Council grants among European funding opportunities).

It is also important to emphasize the role of formal institutional promotion. In this structure, surprisingly, promotion (here, full professorship) is not the central mechanism that legitimizes success. It emerges as a secondary effect of earlier prestige accumulation in the global publication circuit. In field-theory terms, promotion can be interpreted as a shift from institutional capital to symbolic capital with an international reach. This is consistent with observations about the growing importance of measures based on scientific communication. These measures include publications and citations and form the basis of evaluation regimes for scientists and their institutions (Gingras, 2016; Moed, 2005; Wilsdon et al., 2015).

The implications for theories of inequality in science are straightforward. Inequalities do not result only from differences in productivity but also from structural coupling between different dimensions of recognition. In this coupling, the central hub (global publication prestige) reinforces capital accumulation in other areas. Success is therefore relational: the value of an achievement depends on its position in the prestige architecture, not only on its inherent properties (Borgatti et al., 2018; Bourdieu, 1993; Kwiek, 2021). At the same time, pressure for measurable forms of success can lead to strategic adaptation and evaluation pathologies on the part of institutions and individual scientists (Biagioli & Lippman, 2020; Burrows, 2012; de Rijcke et al., 2016; Fochler & de Rijcke, 2017; Muller, 2018). From this perspective, the dominance of global publication-based prestige can consolidate the advantages of actors already located in the center, consistently with classic sociology of science and newer critiques of metric regimes (Cole & Cole, 1973; Hicks et al., 2015; Merton, 1968; Wilsdon et al., 2015).

Academic success should not be interpreted as having a one-dimensional scale. It should also not be interpreted as a set of equivalent and independent indicators. The dependency structure between dimensions of recognition points to an ordered relational configuration in which some elements act as central integrating hubs. Other elements remain peripheral and have lower convertibility (Bourdieu, 1988, 1993; Schott, 1998).

In systems that are strongly integrated into the global circulation of knowledge, one can expect one dimension to have clearly higher centrality than others. This dimension will generate cumulative coupling effects. It will therefore increase the probability of success in other domains. Global capital should have higher convertibility than local capital. Moving to the highest level of recognition should be threshold-based rather than linear. The structure of the success hierarchy should be relatively stable across groups of scientists within the same system. However, the intensity of ratings may vary. It may depend on institutional pressure and indicator-based regimes (de Rijcke et al., 2016; Fochler & de Rijcke, 2017; Wilsdon et al., 2015).

Ultimately, academic success is not a simple sum of publications, citations, or awards. It is the result of being integrated into a prestige architecture. This architecture is network-integrated, hierarchically ordered, and threshold-selective. Understanding how it works helps explain contemporary mechanisms of stratification in science. It also helps describe the tension between global and local definitions of research value. This is especially important in semi-peripheral systems. In such systems, local legitimation increasingly depends on global criteria of recognition (Hazelkorn, 2015; Marginson, 2022; Schott, 1998).

## 6. Conclusions and implications

Our results have important theoretical implications. They show the value of a relational and network-based approach to success in science and confirm an asymmetry in capital convertibility: global capital (especially publications in top journals) is more convertible into further advantages in science than local capital (such as locally awarded full professorships or publications in national journals). The threshold nature of the highest rating of success highlights that change between the very good rating and the top rating is qualitative rather than merely quantitative. Presence at the top does not follow only from having more publications—it requires entering the core of prestige in its various dimensions.

System-level implications are also important. The globalization of science (Kwiek, 2023) imposes a homogenization of national definitions of success and introduces its single, global meaning. Across countries and institutions, scientists increasingly operate with a shared sense of success. In this homogenized view of success, global visibility is dominant. Policies that strengthen internationalization and publication selectivity fit this global architecture very well. However, they may also weaken local knowledge circuits, especially in social sciences and humanities.

Overall, academic success is neither a set of equivalent indicators nor a linear scale of achievements. It is an ordered, relational, and threshold-based prestige architecture. In this architecture, some dimensions are central, integrating various forms of recognition. Other dimensions of success are peripheral and less convertible. Success is therefore not only about achievements but also about their actual position in the structure of relationships.

The core of the system in the Polish case is publishing in top international journals. This dimension has the highest centrality and the highest convertibility into other forms of recognition. Therefore, success is increasingly defined by a position in the global circulation of knowledge. Success emerges from our research as less dependent on formal institutional positions and local prestige channels, including local scientific journals. The most important component of this architecture is global publication visibility.

## Acknowledgments

We gratefully acknowledge the assistance of the International Center for the Studies of Research (ICSR) Lab, with particular gratitude to Kristy James and Alick Bird. We gratefully acknowledge the support of Dr. Łukasz Szymula with Scopus data acquisition and integration.

## Author contributions

Marek Kwiek: Conceptualization, Data curation, Formal analysis, Investigation, Methodology, Resources, Software, Validation, Writing—original draft, Writing—review & editing. Wojciech Roszka: Conceptualization, Data curation, Formal analysis, Investigation, Methodology, Software, Validation, Visualization, Writing—original draft, Writing—review & editing.

## Competing interests

The authors have no competing interests.

## Funding information

We gratefully acknowledge the support provided by the Ministry of Science (NDS grant no. NdS-II/SP/0010/2023/01).

## Data availability

We used data from Scopus, a proprietary scientometric database. For proprietary reasons, data from Scopus received through collaboration with the ICSR Lab (Elsevier) cannot be made openly available.

# Appendix

Table 1A. Robustness checks

| term | Binary GLMM (==5) | | | | Binary GLMM (>=4) | | | |
|---|---|---|---|---|---|---|---|---|
| | OR | OR_low | OR_high | p.value | OR | OR_low | OR_high | p.value |
| (Intercept) | 0.592 | 0.461 | 0.760 | 0.000 | 2.690 | 2.176 | 3.324 | 0.000 |
| Academic age | 1.017 | 0.956 | 1.081 | 0.599 | 0.958 | 0.909 | 1.008 | 0.100 |
| International collaboration rate | 1.046 | 1.007 | 1.086 | 0.020 | 1.043 | 1.010 | 1.076 | 0.010 |
| Discipline: Engineering & Technology (ENGTECH) | 1.059 | 0.943 | 1.190 | 0.334 | 0.982 | 0.891 | 1.082 | 0.711 |
| Discipline: Humanities (HUM) | 0.986 | 0.851 | 1.142 | 0.847 | 0.860 | 0.760 | 0.972 | 0.016 |
| Discipline: Medical Sciences (MED) | 1.343 | 1.178 | 1.530 | 0.000 | 1.239 | 1.109 | 1.384 | 0.000 |
| Discipline: Natural Sciences (NATSCI) | 1.011 | 0.898 | 1.138 | 0.855 | 0.903 | 0.817 | 0.997 | 0.043 |
| Gender: male | 0.529 | 0.492 | 0.569 | 0.000 | 0.651 | 0.613 | 0.692 | 0.000 |
| Institution: IDUB university | 0.973 | 0.830 | 1.141 | 0.740 | 1.022 | 0.894 | 1.167 | 0.752 |
| Institution: non-IDUB university | 0.991 | 0.849 | 1.158 | 0.912 | 1.023 | 0.899 | 1.165 | 0.731 |
| Very high institutional pressure (Q37_5 = 5) | 2.160 | 1.854 | 2.515 | 0.000 | 1.352 | 1.186 | 1.542 | 0.000 |
| Very high intrinsic motivation (Q37_4 = 5) | 1.865 | 1.738 | 2.001 | 0.000 | 1.160 | 1.093 | 1.230 | 0.000 |
| Total number of articles | 1.028 | 0.978 | 1.081 | 0.284 | 1.055 | 1.011 | 1.102 | 0.014 |
| Success dimension: Elected membership in associations, academies etc. | 0.192 | 0.176 | 0.211 | 0.000 | 0.264 | 0.246 | 0.283 | 0.000 |
| Success dimension: Publications in top international academic journals | 6.358 | 5.882 | 6.872 | 0.000 | 7.597 | 6.939 | 8.318 | 0.000 |
| Success dimension: Publications in top Polish academic journals | 0.207 | 0.189 | 0.227 | 0.000 | 0.198 | 0.184 | 0.213 | 0.000 |
| Success dimension: Research funding / grants obtained | 0.983 | 0.912 | 1.061 | 0.663 | 1.360 | 1.266 | 1.460 | 0.000 |
| Success dimension: Polish distinctions and awards | 0.219 | 0.200 | 0.239 | 0.000 | 0.348 | 0.324 | 0.373 | 0.000 |
| Success dimension: Broad international contacts | 1.283 | 1.190 | 1.382 | 0.000 | 2.130 | 1.978 | 2.293 | 0.000 |
| Success dimension: High number of citations | 1.692 | 1.570 | 1.822 | 0.000 | 2.274 | 2.111 | 2.449 | 0.000 |
| Success dimension: Plenary talks at international conferences | 0.468 | 0.432 | 0.507 | 0.000 | 0.658 | 0.614 | 0.705 | 0.000 |
| Success dimension: Employment at a prestigious institution | 0.485 | 0.448 | 0.525 | 0.000 | 0.760 | 0.709 | 0.815 | 0.000 |
| Academic position: Assistant Professor | 0.618 | 0.540 | 0.708 | 0.000 | 0.734 | 0.654 | 0.823 | 0.000 |
| Academic position: Associate Professor | 0.712 | 0.632 | 0.801 | 0.000 | 0.803 | 0.726 | 0.887 | 0.000 |
| Median article prestige | 1.058 | 1.018 | 1.100 | 0.004 | 1.021 | 0.989 | 1.055 | 0.197 |
| Language of research: English | 1.075 | 0.992 | 1.164 | 0.077 | 0.973 | 0.910 | 1.040 | 0.419 |
| Top performers: Rest | 0.897 | 0.786 | 1.025 | 0.110 | 0.951 | 0.850 | 1.064 | 0.385 |
| Biological age | 0.947 | 0.894 | 1.003 | 0.064 | 1.065 | 1.015 | 1.117 | 0.010 |